# Differential Imaging Forensics: A Feasibility Study

Aurélien Bourquard
Massachusetts Institute of Technology, USA
Universidad Politécnica de Madrid and CIBER-BBN, Spain
aurelien@mit.edu

Jeff Yan
University of Strathclyde, UK
jeff.yan@strath.ac.uk

## ABSTRACT

We motivate and develop a new line of digital forensics. In the meanwhile, we propose a novel approach to photographer identification, a rarely explored authorship attribution problem. We report a proof-of-concept study, which shows the feasibility of our method. Our contributions include a new forensic method for photographer de-anonymization and revealing a novel privacy threat which had been ignored before. The success of our creation builds on top of a new optical side-channel which we have discovered, as well as on how to exploit it effectively. We also make the first attempt to bridge side channels and inverse problems, two fields that appear to be completely isolated from each other but have deep connections.

## 1 INTRODUCTION

We consider the following research problem: given a *single* photo, how to determine who was the cameraman? This is in general a hard problem, except for selfies, or if the photographer's shadow became visible in the photo, or if her image was captured by a reflective object in the photo, for example, by a subject's eyes.

This problem is interesting to intelligence agencies. For example, a photo of a secret military facility in Russia can be valuable to the Central Intelligence Agency (CIA) of USA. However, when the photo gets leaked by a mole inside the CIA, Russia's anti-spy operatives would be keen to work out who took the photo in the first place.

The problem is interesting to law enforcement agencies, too. For example, when the Scotland Yard are tipped off by a photo from an anonymous source that offers clues to a criminal investigation, it is likely to gain further information to accelerate their investigation by identifying the person behind the camera.

Moreover, the problem is also interesting to privacy researchers. The answer to the research question will not only likely provide novel privacy intrusion methods by de-anonymising a photographer of any concerned photo on the Internet, but also motivate novel research for protecting photographers' anonymity.

Not all photographers care that it is public knowledge that some photos are taken by them. However, in some circumstances, some photographers do have to care.

From the forensic perspective, a technique that does not identify the photographer 100% of the time can still be practically useful, since it will narrow down suspects to a small number. Complemented with other means such as surveillance, it is highly likely for intelligence agencies or law enforcements to pin down the concerned photographer accurately.

We first review related work in the literature, and we show that existing approaches do not resolve the research question we are asking. Then, we propose a new approach, and demonstrate its feasibility by a proof-of-concept but realistic simulation study. Specifically, our method entails a new optical side-channel and leverages its characteristics for the estimation of distinctive physical features of a photographer. Our method is applicable to both digital and film photography, in theory.

## 2 RELATED WORK

**Visual stylometry**. Artists like Claude Monet and Vincent van Gogh demonstrate distinctive styles in their paintings. In the past hundreds of years, people relied on stylistic analysis to tell apart genuine fine art from fakes. It became an emerging research area in recent years to apply signal processing and machine learning methods to analyse painting images for artist identification [1, 2].

Similarly, some photographers display peculiar styles in the photos they produce. For example, widely regarded as one of the best portrait photographers of all time, Yousuf Karsh has been known for distinctive features in his portraits due to lighting, composition, and posture. Ernst Haas showed a distinctive personal style in his impressionist colour photography, too. Therefore, it is a natural extension to develop photographer identification methods from painting artist identification.

However, a training set of photos, usually of a large size, is needed for each concerned photographer to make machine-learning methods to work. This approach will hardly work if the given photo is the only available one taken by a suspect photographer, since it is impossible to collect a training set of photos for the photographer. On the other hand, if a photographer's style is not sufficiently sophisticated, it is easy for somebody else to emulate. This can be exploited to fool machine learning algorithms, and to frame a photographer.

**Camera fingerprint** [3,4]. CCD or CMOS imaging sensors are a digital camera's heart. Due to sensor design and imperfections of the sensor manufacturing process, systematic artefacts (usually known as sensor pattern noises) form an equivalent of a digital fingerprint that can identify a camera. Such fingerprints are intrinsically embedded in each image and video clip created by a digital camera. Forensic applications of camera fingerprints include 1) source camera identification (which camera was used to produce this image?), and 2) device linking (were two images produced by the same camera?).

Camera fingerprint, in theory, can link a photo to a specific camera, if a reference fingerprint can be established for the camera, e.g. when the camera is physically accessible, or a set of photos taken by the same camera is otherwise available. However, camera fingerprint does not link a photo to a specific user of the camera. This is an issue when the same camera has been used by many people. Moreover, camera fingerprint can be easily removed from each photo, entirely disabling its forensic applications. On the other hand, the camera-fingerprint technique has been developed for digital cameras, and it does not work for traditional film photography.

**Image metadata** has a limited forensic application. For example, it can link a digital image to a camera model at most, but not to a specific camera, let alone a photographer. On the other hand, film photography does not produce any such metadata.

None of the methods discussed above provide a solution that properly addresses our research question.

## 3 A NOVEL METHOD

When a scene is photographed, a photographer's body often deflects light (via reflection and refraction) into the scene, leaving an impact on a photo created thereafter. Our hypothesis is that light rays deflected by a photographer into the photo that she is taking will give away some physical characteristics of herself.

Elaborating on the earlier results we reported in [5], we refine our research problem as follows. We are given a photo $P_1$, which was taken of a scene by a photographer at will, i.e. its acquisition is non-controlled; and our task is to work out who took the photo. We have access to the same physical scene, and we take a photo $P_2$ which is similar to $P_1$, while all acquisition parameters are reproduced in a controlled manner to be the same as used for producing $P_1$, except that the photographer is absent. Our research questions are: 1). What differences in $P_1$ and $P_2$ can be exploited to deduce the photographer's physical characteristics? 2). Under what conditions will the above measurement work for that purpose?

We choose to answer these questions via realistic simulation, rather than an empirical lab study, since the latter involves with experiments that are more expensive and more sophisticated to set up. Specifically, we use photon mapping, a well-established ray tracing technique, to conduct a proof-of-concept feasibility study. Photon mapping realistically simulates the interaction of light with different objects. In this approach, light rays from a light source and rays from a camera are traced independently until some termination criterion is met. Then, they are connected in a second step to produce a radiance value.

The dual-acquisition framework proposed above is applied here to synthetically generated photos. Otherwise, this framework is similar to the paradigm that we named as *differential imaging forensics* in [6]. Following acquisition, the photographer's physical characteristics must be deduced from the available photos $P_1$ and $P_2$, which cannot be done otherwise because the photographer is not directly visible in the scene. In general, estimating parameters that cannot be directly observed in the available data amounts to solving a so-called *inverse problem*, which is typically hard to do due to either lack of information or the complexity of the relationship between the parameters and the corresponding acquired data [7]. In what follows, we propose a set of proof-of-principle methods that could be used to deduce the photographer's key physical characteristics.

### 3.1 Experimental Design

We use the popular POVRay software[1] for scene definition and rendering, as well as photon mapping.

#### 3.1.1 Scene Definition.

Fig. 1 illustrates the scene that we use for this study, as viewed from the camera. The ground consists of a brown surface of RGB colour (0.80, 0.55, 0.35), the value of each colour channel being normalised between 0 and 1. The camera capturing the scene is placed 1.5m above the ground, and 2m away from a dark wall that is modelled as a non-reflective rectangular object of 1m width and 1.9m height. This wall casts a shadow on the floor, because of a light source 3m high and 1m behind the wall. The camera's angle of view is 90°, which many lenses can achieve in photography, and the camera is oriented towards the corner formed by the floor and the wall.

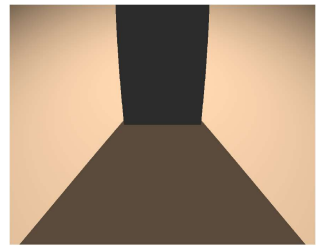

Fig. 1: Scene as viewed from the camera, without the presence of the photographer, as defined in Sect. 3.1.1. The wall (black) and its shadow on the floor (brown) are visible in the picture.

The ground and wall surfaces are flat and modelled with ambient-light and diffuse-light coefficients of 0.1 and 0.9, respectively. The resolution of the rendered scene is 1600x900 pixels. The bit depth is 16 bits per colour channel, in RGB format; this allows to minimise numerical errors.

The photos $P_1$ and $P_2$ are acquisitions of the underlying 3D scene described and shown above, from the point of view of the camera, taken with or without the photographer's presence, respectively.

---
[1] http://www.povray.org

Accordingly, every picture is a rendered version of the scene produced by POVRay.

### 3.1.2 Photographer.

When present, the photographer faces the wall in the scene and stands together with the camera. The photographer's jacket is modelled as a reflective rectangular object whose width and height are free parameters. The jacket exhibits surface irregularities in form of bumps, whose characteristic widths parallel to the surface is 30cm, and whose depths normal to the surface is left as a free parameter, as for the case of the jacket colour. Accordingly, the jacket material reflects light from the light source onto the floor of the scene.

The irregularities of the jacket surface are modelled through a 2D map assigning a specific height--as defined perpendicularly to the surface and relative to its ground level--to every surface location. Based on this map, these irregularities are rendered through a so-called *bump-mapping* technique, which, as an approximation, determines how such height alterations (if any) deflect light paths without modifying the object geometry. This allows to accurately account for the surface properties without being detrimental to the computational complexity of the rendering process. Surface irregularities are modelled as a set of randomly located bumps, their depth being set by a fixed parameter. The light reflections from the body surface onto the floor and the wall are simulated using photon mapping with a count of $20 \times 10^6$ photons, a figure that is empirically determined to be sufficient to converge to a high-quality scene rendering. Photon mapping is essential to model the effect of reflected light from the complex jacket surface onto the rest of the scene (especially the floor) by simulating trajectories of individual photons emitted from the light source and inferring their distribution accordingly.

Based on the pair of photographs $P_1$ and $P_2$, the parameters that are estimated with our method are the (a) height, (b) width, and (c) colour of the photographer's jacket, where for simplicity the jacket dimensions are assumed to match the photographer's dimensions, and (d) the presence of bumps of distinct depths on the jacket surface. A non-flat surface typically exhibits bumps, whose size and depth normal to the surface may vary; a zero depth is equivalent to a flat surface. In conjunction with colour, the presence and dimensions of bumps characterize the type or class of fabric material used. Indeed, fibres composed of different fabrics are expected to modulate light-reflection properties differently through their surface irregularities. The observed light-brightness distribution on the photo $P_1$ used for estimation is thus expected to vary accordingly, which can be used for estimation.

Each of the parameters (a)-(d) is varied within a certain range and compared to corresponding estimates, using 8 data points in total, and using default values for the other parameters. The jacket width ranges from 0.5m to 1.5m. The jacket height ranges from 1m to 2m. The depth of bumps normal to the jacket surface ranges from 0 to 20cm. Following the RGB convention, the jacket colour ranges from 0 to 1 for the G channel, the other colour channels being left to their default values.

In the 3D scene, the estimates of all parameters associated with geometrical lengths and distances are made in pixels. Accordingly, what is assessed between reference and estimated values is mere proportionality, as opposed to strict equality with the original dimensions defined in meters.

In our study, each parameter of interest is varied and estimated independently while other parameters stay constant at their default values. This allows to infer preliminary yet indicative proof-of-principle results where confounding factors are minimized. The default values for the width and height of the photographer's jacket are set as 0.75m and 1.5m, respectively. The depth of bumps normal to the jacket surface is set as 10cm by default. Following the RGB convention, the jacket colour default is set to (1, 1, 0).

### 3.1.3 Noise levels.

In addition to these specific parameters that characterise the scene in a unique way, we also consider varying acquisition-quality levels through various signal-to-noise ratio (SNR) values for the resulting photographs, taking into consideration, as a representative example, the artificial addition of camera-sensor noise at the pixel level modelled as Gaussian noise.

For every parameter and estimation method described below, we consider several simulated noise levels in rendered scenes, both with and without the photographer, corresponding to distinct SNR levels (defined as the ratio between the energies of the signal and of the noise) for the photographs in decibels (dB). Specifically, we consider cases with 25, 30, 35, 40, 45, 50, and $+\infty$ dB (the noiseless case).

The noiseless case corresponds to an ideal replication of photo acquisition conditions, including the camera being the same model. Other noise levels help to lessen our tight control on the acquired data, by allowing for example a camera different from the one used by the original photographer, and by accounting for the presence of sensor noise.

## 3.2 Results and Discussions

In this section, we report and discuss the results of our estimates compared to the ground-truth free parameters mentioned in the previous section. Note that, except for colour, the estimates that we propose below are not meant to consistently scale with the corresponding ground-truth parameters, though they are expected to monotonically increase with their values.

### 3.2.1 Photographer Body Size.

In view of estimation, the scene difference D is computed as the pixelwise difference between the rendered scene (1) with the photographer present and the rendered scene (2) without the photographer.

Based on this difference map D defined in RGB, the grayscale scene difference $D_G$ is first computed by averaging the 3 channel values pixelwise. To isolate the object profile of interest for inferring its dimensions, we employ thresholding as a means of segmentation. The thresholded scene difference DT is computed as a thresholded version of DG, i.e.,

DT = 1, for pixels above τ times the maximum
   intensity max(DG), and

DT = 0, otherwise.

We empirically set τ=1/20. These steps make size estimation independent of brightness and colour information.

The photographer's size (width and height) is directly estimated based on the light that the jacket body reflects onto the floor. These dimensions are assumed to exactly match those of the floor area impacted by the light reflection. This is merely a coarse approximation, compared to other methods that consider camera parameters and object geometry. However, this estimation approach is sufficient for our proof-of-principle study. Specifically, our estimation is based on DT. We first locate the gravity centre of the pixels in DT s.t. DT=1, and then estimate the width and height as the root-mean-square (RMS) horizontal and vertical distances of all such pixels with respect to the said gravity centre. Specifically, the RMS width and height are defined as

$W_{RMS} = [S^{-1} \Sigma (x - x_0)^2]^{1/2}$,
$H_{RMS} = [S^{-1} \Sigma (y - y_0)^2]^{1/2}$,

with $(x_0, y_0)$ the gravity-centre coordinate, $\Sigma$ the sum over the pixel domain inside which DT=1, and S the total domain surface.

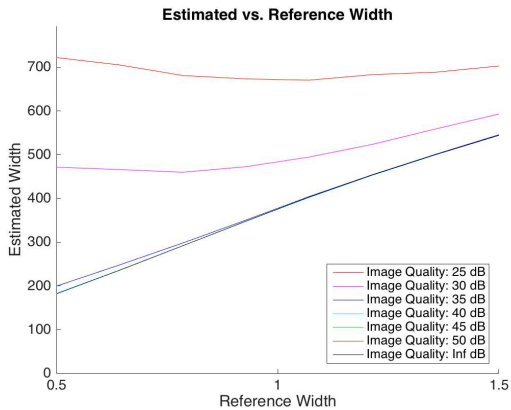

Fig. 2: Estimated vs. reference width in the 0.5~1.5m range

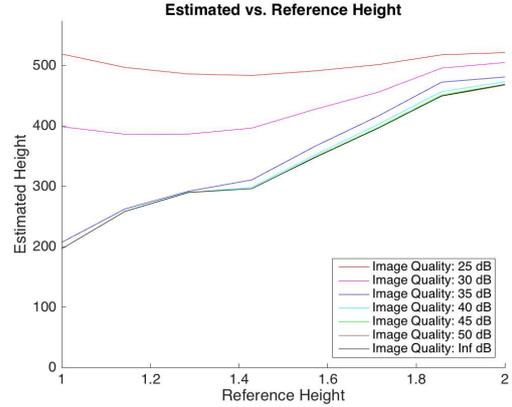

Fig. 3: Estimated vs. reference height in the 1~2m range

Figs. 2 and 3 show our estimation results. The plots show that width and height can be estimated based on the rendered scenes, even though the relationship is not linear. In particular, the increase in the estimated values is monotonic with respect to the corresponding scene-parameter reference values for these geometric features. The estimation starts to break down below a certain SNR level, because noise affects the thresholded scene-difference image and hence the estimation of the jacket dimensions.

Figs. 4 and 5 show rendered scenes corresponding to two extreme heights we consider. As shown in Fig. 5, the maximum-height value corresponds to a case where the reflected photons vertically spread over the whole field of view, with small patches scattered around.

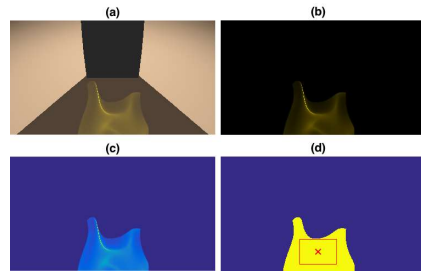

Fig 4: The minimum height case: (a) rendered scene, (b) pixelwise RGB difference (w – w/o photographer), and the difference map's brightness: (c) before & (d) after thresholding.

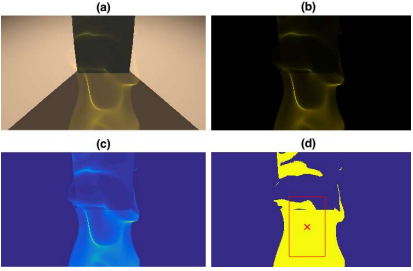

Fig. 5: The maximum height case: (a) rendered scene, (b) pixelwise RGB difference, and the brightness of the difference map: (c) the original; (d) after thresholding.

With proper calibration, width and height estimates can be used by forensic specialists to roughly estimate the photographer's body size, which for instance could serve as a proxy for age.

3.2.2  Fabric Surface Irregularity.

Bumps modify the way the photographer's jacket reflects light onto the rest of the scene by modifying the local surface normal impacted by light rays. Here, we consider varying the *bump depth* B normal to the surface, defined in centimetres and equal to or greater than zero.

When bump depth increases, the angles of light reflections on a given surface patch will become less uniform. Accordingly, the illumination stemming from that light on a plane floor is expected to become less and less spatially uniform in terms of intensity. Accordingly, we propose to estimate bump depth based on the analysis of the horizontal and vertical spatial derivatives the pixel-brightness values, i.e., based on the magnitude of the spatial gradient. The gradient essentially captures how fast the brightness varies between adjacent pixels, thus highlighting transitions and irregularities in the image under consideration. These gradient values are further filtered by a Gaussian filter (i.e., an averaging filter based on the Gaussian function) whose effect is to smooth out the noise. Hence, high frequencies associated with noise are decreased due to Gaussian smoothing, while brightness irregularities and transitions associated with non-zero bump depth are suitably captured by the directional derivatives in the gradient.

Specifically, based on the non-thresholded grayscale scene difference DG, we estimate bump depth viewed from the rendered-scene picture considering the region of interest (ROI) R corresponding to the estimated width and height of the object. We then consider the gradient image RG of a Gaussian-smoothed version of this area, considering a 2D Gaussian filter with standard deviation of 5 pixels, and a gradient filter based on the finite-difference first derivative [1, 0, -1]. The resulting bumpiness estimation is finally defined as:

bumpiness = *mean*(RG) / *mean*(R),

i.e., the spatial average of the gradient value within the ROI normalized by the average ROI intensity.

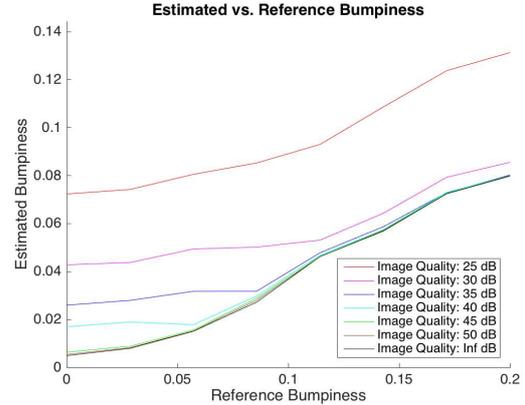

Fig. 6: Estimated vs. reference bumpiness in the 0~20cm range

Estimation results (Fig. 6) show that the estimated bump depth value increases with the corresponding scene parameter. Noise also affects the estimation because it increases the perceived surface irregularity viewed from the rendered scene, even though Gaussian filtering was used to mitigate the effect.

As bump depth increases, the perceived irregularities on the visible jacket-related light impacting the floor also increases, which is especially noticeable when comparing two extreme cases, i.e. a bump depth of zero (i.e., a flat surface) and the maximum bumpiness considered in the range of interest, as illustrated in Figs. 7 and 8.

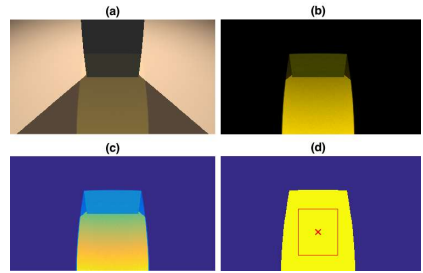

Fig. 7: The minimum bumpiness (i.e. a flat surface) case: (a) rendered scene, (b) pixelwise RGB difference, and the difference map's brightness: (c) before & (d) after thresholding.

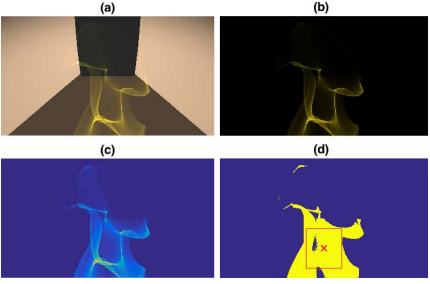

Fig 8: The maximum bumpiness case: (a) rendered scene, (b) pixel wise RGB difference, and the brightness of the difference map: (c) the original; (d) after thresholding.

From a forensic perspective, the bump-depth estimate could provide information on fabric materials of the clothes the photographer was wearing. It can be hard to pinpoint exact materials, but such an estimate has the potential to classify cloth materials into several categories.

### 3.2.3 Jacket Colour.

The jacket colour is estimated based on the RGB rendered-scene difference within the ROI described above. Specifically, every colour-channel scalar value is estimated as the spatial mean within the ROI.

In general, the luminance of a measured colour depends on several factors other than the material under consideration, such as reflectance and camera-sensor parameters. To mitigate this fact, we choose to estimate an objective specification of the quality of colour regardless of its luminance. Specifically, we compare the three normalized channels in the scene vs. in the estimation which essentially contain the chromaticity information, with the normalized channels R', G', B' being defined as follows.

R'=R/(R+G+B);
G'=G/(R+G+B);
B'=B/(R+G+B).

Fig. 9 shows our estimation results. The relationship between the colour estimates and its corresponding values in the scene closely follows a linear trend, even under noise. This is because, unlike the original RGB channels, chromaticity information is robust to potential changes in light reflection and illumination on the floor, and robust to potential ROI misestimation or overestimation due to noise.

From a forensic perspective, the colour estimate could provide information on the type of jacket fabric material or dye, each material or dye being associated with a specific colour. This estimate is also expected to be robust to illumination since it does not depend on luminance.

## 4 EXTENSIONS

When a set of photos taken by the same photographer is available, with our method, each might contribute a snapshot of her body profile. A reconstruction algorithm can be envisaged to build a full 2D or 3D profile for the photographer.

Further investigations are also warranted to determine when the proposed method precisely starts to break, with regard to the level of noise; and to determine what is the required exposure time and scene characteristics to reach a target SNR. This will allow determining what real scenarios and settings are conducive to successful acquisition and parameter estimation, similar to the results that we report for high enough SNR values in Section 3.2.

Finally, practical experiments reproducing our simulations could be carried out using a camera and a mannequin in a controlled environment that includes precise object and light source positioning, similar to those in [6].

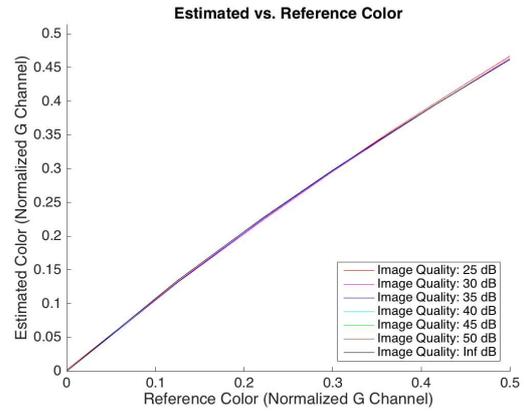

Fig. 9: Estimated vs. reference values in the normalised (dimensionless) 0~0.5 range for the green-colour channel.

## 5 KEY INSIGHTS

Our key insight is that, when a scene is photographed, a photographer's body is likely to deflect light (e.g., via reflection or refraction) into the scene, leaving impacts on a photo created thereafter. By setting up the same photographic scenario and acquisition conditions for the same scene, but without a reflective body behind the camera, we can produce a second photo. The difference between these two photos can then be computed with image processing methods. It gives crucial clues for deducing specific physical characteristics of the photographer. In a sense, the photographer gives away her own presence via an optical side channel which was not investigated in the literature before.

## 6 BRIDGING SIDE CHANNELS AND INVERSE PROBLEMS

The concept of a side channel originated from cryptanalysis and systems security. Security people and cryptologists care about information accidentally leaked from side channels, as well as proper countermeasures. On the other hand, the research of inverse problems has deep and historic roots in mathematics. The main source of inverse problems are science and engineering. These problems have pushed not only the development of mathematical theories and tools, but also scientific and technological innovations in a wide range of disciplines, including astronomy, geophysics, biology, medical imaging, optics, computer vision, among others. However, side channels and inverse problems appear to be two fields that are completely isolated from each other, and few researchers in one community even know the existence of the other.

## 6.1 Side Channels vs. Inverse Problems

Our work reveals a close link between side channels and inverse problems, perhaps for the first time. In a side channel, information leaks accidentally via some medium or mechanism that was not designed or intended for communication. Often, a direct measurement of the output from a side channel does not immediately give away the information leaked. Instead, the direct output measurement is akin to metadata, from which attackers deduce the leaked information.

Similarly, in our experiments, the channel output consists of images of a scene where the photographer is not directly visible, and from which the information of interest, such as the photographer's physical parameters, is inferred from the methods presented above.

Whenever the leaked information is not directly visible, its estimation from the available data can be understood as the resolution of an inverse problem. Indeed, inverse problems consist in the estimation of parameters of interest that produced a set of available observations, thus inverting the causal chain, whenever the link between both is non-trivial. In general, the amount of information that is leaked via the side channel could give an upper bound on how accurately such an inverse problem can be resolved. The estimation of such an upper bound in our setting remains a topic of future research.

We use SonarSnoop [8], which one of us conceived, as another example to illustrate the link between side channels and inverse problems. From a security perspective, SonarSnoop turns a smartphone into a sonar system, where built-in speakers emit human-inaudible acoustic waves and where microphones capture the echoes thereof. When people move fingers around to draw Android-unlock patterns, their fingers echo acoustic signals emitted from the speakers, and the echoes detected by the microphones will be used to reconstruct the drawings. This way, an acoustic side-channel was established to steal a victim's authentication secret.

On the other hand, inferring the drawing (which was unobservable to an attacker) based on the available data, i.e., the acoustic signals detected by the sonar system, clearly constitutes an inverse problem. Even without the knowledge of [8], an inverse-problem expert would formulate the problem and potential solutions thereof as follows. What is the drawing that best explains the available acoustic data? To answer that question and resolve the corresponding inverse problem, one may additionally use prior knowledge on what drawings make most sense and/or are most likely, to mitigate the possibility that several solutions may lead to the same observations, e.g. in the case of incomplete information. It turns out that all these steps exactly match what was done in [8], despite the use of different terminologies.

In general, every side channel implies (or involves) an inverse problem, but not vice versa.

Examples of inverse problems involving no side channels can be many: whenever the measurements that are used are intentionally (as opposed to accidentally) created by the device of interest, yet not trivially related to the information to be retrieved in its final form. We give computer tomography (CT) and Magnetic Resonance Imaging (MRI) as two such examples, both classical achievements of inverse problems. The fundamental inverse problem for CT is a mathematical one, namely, determining the density $\rho$ of a particular location in a human body for a given Radon transform $R\rho$, as measured by the intensity attenuation of thin X-ray beamed into the human body. Inverse Radon transform is also fundamental to MRI, which is however based on some physics different from those for CT.

It is interesting to note that, in some instances, the inverse problem implied by a side channel may be relatively straightforward. For example, Markus Kuhn demonstrated a classical optical side-channel, where the information displayed on a computer monitor could be reconstructed remotely by decoding the light scattered from the face or shirt of a user sitting in front of the computer [9]. A sophisticated attack was required to successfully exploit this side channel. However, its key insight was the fact that the whole screen information was available as a time-resolved signal, rather than solving a complex inverse problem.

## 6.2 How do They Inform Each Other?

In unclassified worlds, side channels are a young field, with a history of less than forty years. However, inverse problems have been studied for more than a century. Not all inverse problems involved in side channels are straightforward to solve. For example, SonarSnoop [8] involved a rather complex inverse problem. Therefore, it is likely that side channels will become a new playground for inverse problem researchers. Moreover, some interesting questions can be asked along these lines: How will both fields inform each other? What existing methodologies and techniques could be transferred from one field to the other in a way that is beneficial? Can valuable lessons be learned, and new research directions be defined?

A problem is 'inverse' because it starts with the observable effects to calculate or infer the causes, such as determining causal factors and unknown parameters from a set of observations of a system of interest. It is the inverse of a forward (or direct) problem, which starts with the causes and then deduces or calculates the effects, such as modelling a system from known parameters.

A fundamental challenge in inverse problems is they are typically ill posed in terms of the solution's existence, uniqueness, and stability, whereas their corresponding forward problems may be well posed in all these regards. The property of stability in this context means that a solution depends continuously on the measurements (i.e. the data) available. Therefore, a problem lacks stability if adding or removing data implies a radically different solution. If a computed solution lacks stability, it will simply depart from the true solution.

To solve challenging inverse problems, mathematics has been applied to accurately describe the forward (physical) model as well as assumptions on the solution, if any. Specific theories, approaches, and key insights have led to well-posed problem formulations and solutions. For instance, sound statistical modelling allows reducing the dimensionality of the parameter spaces and producing accurate solutions, and efficient algorithms also allow maximizing computational efficiency.

We believe that problem-formalisation strategies, theoretical models, mathematical techniques, algorithms, and concepts developed in the context of inverse problems could benefit and inspire future research of side channels in multiple ways.

First, the inverse-problem literature might inspire researchers to build a similar scientific foundation and framework for side channels, including a more rigorous and less ad-hoc problem formalisation and structure, along with a range of tools and concepts. This could happen just in the same way as the relevant literature in applied mathematics benefited the research on inverse problems once computational power became sufficient to compute sophisticated models and high-dimensional data.

Second, it also helps to properly navigate between the languages used in both fields. This will, for instance, help to identify similarities and differences, to clarify misconceptions, and to unify terminologies.

For example, *information*, which is the set of relevant parameters approximated by the solution to the inverse problem, vs. *measurements*, which are the physically leaked raw-data input of the inverse problem and which can contain various amount of useful information.

In a unified language that is understandable for both fields, blocking a side-channel attack essentially amounts to making the corresponding inverse problem unsolvable, intractable, or at least harder to model, or harder to compute efficiently. Accordingly, there are the following three scenarios: (a) prove that the inverse problem becomes impossible to solve by getting rid of the information that is present in the measurements, in such a way that the analysed measurements contain no useful information; (b) make the inverse problem much harder to model mathematically or solve computationally; (c) get rid of the leakage (e.g. physically) so that there are no measurements to exploit whatsoever (regardless of whether the said measurements contain meaningful information or not). Adding random perturbations such as noise is a classical mechanism that makes an inverse problem unsolvable or harder to model.

Third, the literature on inverse problems contains useful theoretical results, including reconstruction guarantees for several types of problem structure, lower bounds on the reconstruction error (Cramér-Rao bounds)[2], and the extent to which the recovery is affected by noise or other non-idealities (which amounts to mitigating side-channel attacks in the context of security and cryptanalysis). Such results could inform on how to best characterise various side channels (including acoustic, electromagnetic and optical ones), and how to best design and evaluate their countermeasures.

For example, in a systematic study of acoustic side channels which we recently conducted, we noticed a peculiar but interesting phenomenon: most of the literature, if not all, focused on attacks, yet only discussed possible countermeasures briefly. Only a few works implemented the countermeasures and evaluated their performance and effectiveness. Even fewer compared different countermeasures, or developed metrics for that purpose, or investigated the best possible countermeasures in a comparable way, either in theory or in practice. We believe these areas of research can benefit much from the skillsets of inverse problems and lead to useful results.

Moreover, some security experts may have been more eager to discover new side channels, which amounts to proving the existence of a solution for the corresponding inverse problem, than to investigate the two related properties, uniqueness and stability. Therefore, looking into these properties, as studied from the perspective of inverse problems, will essentially give security researchers a new lens for examining side channels, as well as their countermeasures.

In retrospect, this is the case for SonarSnoop, for instance. Indeed, once we had established a novel and effective side-channel attack, we did not push further to carefully characterise the side channel, for two reasons. First, that was and still is the common practice in security research, as it is often already challenging enough to establish and exploit a novel and sophisticated side-channel. Second, nobody realised at all at the time that the research on a side channel could be formulated from the perspective of inverse problems. In view of these, it is now obvious that much broader and deeper research can be conducted on SonarSnoop to understand both the side channel itself and possible countermeasures.

For example, examining the stability property alone warrants interesting research, both in the specific case of SonarSnoop and in general. Specifically, how will the side channel be impacted if less, or more, measurement data are collected for experiments? How much measurement data is necessary for the side channel to be stable, in such a way that the retrieved information depends continuously on that data (as opposed to varying abruptly across nearly similar datasets)? Could specific countermeasures, such as adding some type of physical disturbance or interference, influence the observed output from the side channel in such a way that stability decreases? Answers to these questions could allow better optimising side-channel countermeasures, quantifying their efficiency, and providing a robust framework to compare them in a systematic and rigorous manner.

---

[2]These reconstruction guarantees are often only tied to the forward model mapping the relationship between the information of interest and measurements, in the sense that they do not depend on any specific algorithm or solution used.

Finally, we were also intrigued by possible connections between the optimality[3] of a side channel in a given scenario and the uniqueness and stability of the solution to the corresponding inverse problem. In some cases, it appears that the latter indeed implies an optimal side channel. However, in many other scenarios, whether such a connection holds or not has no straightforward answers. Instead, these will be an interesting topic for future research.

# 7 CONCLUSIONS

We have proposed a dual-acquisition framework to address the hard problem of determining who is the photographer, given a single photo. For a given photo of interest, we take a second photo of the same scene under the same acquisition conditions but without a reflective body behind the camera. Then, we compute the difference between these two photos with image processing methods. Finally, we retrieve useful information on the photographer based on specific proof-of-principle parameter-estimation methods. Specifically, we have used photon mapping, a well-established ray tracing technique, to show that our approach in theory can contribute to estimate the photographer's physical characteristics such as body size, clothes colour, and fabric materials.

Overall, our results suggest the possibility to narrow down the search in the task of photographer identification based on a novel set of visual clues in a scene. It is important future work to establish the validity of our method and its limitations through empirical studies, and to roll it out from laboratories to the real world.

It is worthwhile to note that some practical experiments based on a similar acquisition paradigm have already been successfully carried out in [6]. It appears to be the most promising to combine the parameter estimation approach introduced in this paper with the type of setup proposed in [6].

We believe that our forensic approach is applicable to both digital images and traditional print photos since it is a type of photo content analysis. We note that there are methods available for probing who is possibly the photographer in traditional photography. For example, it is possible to analyse a print photo to identify processing chemicals used, such as developer and fixer, and then link them to a particular darkroom or film-processing laboratory. Sometimes, it is also possible to examine human fingerprints that are detectable on the print. Therefore, for film photography, our method is both an alternative approach, and a complementary one.

Our main contributions are threefold. First, we have proposed a novel forensic method for de-anonymising a photographer, with applications in both forensics and privacy intrusion.

Second, we have identified a novel optical side channel, and proposed a methodology to exploit it successfully. These are foundational to our new forensic method.

Third, we have made the first attempt to bridge side channels and inverse problems. Although it may be a small step forward at this stage, it is perhaps the start of an aspiration that will grow in the future. We believe that this bridge has the potential to foster cross-field collaboration and inspire several new research directions, for example, building a more rigorous and effective scientific foundation for side channel research, and encouraging the possibility for ideas and techniques originated in one field to enjoy a wider applicability than was previously anticipated.

## Postscript

In a sense, this is an extended version of our earlier work [5]. More than that in a significant way, however. In view of the empirical evidence that we reported in [6], one can be more assertive regarding the prospect of differential imaging forensics than at the early stage when we were still pondering the possibility of venturing out "towards some new forensics".

We have neither intention nor interest to produce minimal publishable units. Instead, what we are interested in is a public audit trail, documenting the conception and development of differential imaging forensics.

---

[3] By optimality, we mean that the maximum amount of information that can in theory be leaked from a side channel is extracted.